\documentclass[aps,prl,twocolumn,groupedaddress,showpacs]{revtex4}

\usepackage{graphicx,subfigure}

\begin{document}

\title{Nonclassicality of Thermal Radiation}

\author{Lars M. Johansen}

\affiliation{Department of Technology, Buskerud University
College, P.O. Box 251, N-3601 Kongsberg, Norway}
\email{lars.m.johansen@hibu.no}
\date{\today}

\begin{abstract}

It is demonstrated that thermal radiation of small occupation
number is strongly nonclassical. This includes most forms of
naturally occurring radiation. Nonclassicality can be observed as
a negative weak value of a positive observable. It is related to
negative values of the Margenau-Hill quasi-probability
distribution.

\end{abstract}

\pacs{03.65.Ta, 44.40.+a, 06.30.-k} \keywords{Thermal states,
nonclassicality, weak values, weak measurements, Margenau-Hill
distribution, P-distribution} \maketitle

The very signature of quantum mechanics, Planck's constant
$\hbar$, was measured for the first time in 1900 in blackbody
radiation \cite{Planck-TheoGeseEnerNorm:00}. In order to derive
the spectrum for blackbody radiation, Planck had to assume that
the energy of the radiation field could only assume discrete
values. This was the first indication that the classical
electromagnetic theory of radiation did not provide a fully
satisfactory model of nature. However, Planck himself maintained
that the radiation field could be described by classical
electromagnetism. For him, it was the emission from the material
oscillators that was quantized \cite{Planck-TheoHeatRadi:59}. More
recent work has confirmed that the blackbody spectrum can be
explained in terms of stochastic electromagnetism
\cite{Boyer-DeriBlacRadiSpec:69}.

The invention of the laser around 1960 sparked the development of
a whole new are of physics, quantum optics. It was found that
coherent states, a non-thermal state of light, was a natural
building block in a quantum theory of radiation
\cite{Glauber-CoheIncoStatRadi:63}. Glauber and Sudarshan
demonstrated that any density matrix could be expanded diagonally
in terms of coherent states
\cite{Glauber-CoheIncoStatRadi:63,Sudarshan-EquiSemiQuanMech:63},
\begin{equation}\label{eq:pdensity}
    \hat{\rho} = \int d^2 \alpha P(\alpha) \mid \alpha \rangle
    \langle \alpha \mid.
\end{equation}
The weight function $P(\alpha)$ is known as the Glauber-Sudarshan
$P$-distribution. Glauber also proposed that nonclassical states,
i.e., states that cannot be modelled by a classical theory, are
those for which the $P$-distribution fails to be a probability
distribution \cite{Glauber-CoheIncoStatRadi:63,%
Titulaer+Glauber-CorrFuncCoheFiel:65}. Put differently, any state
which cannot be expressed as a classical mixture of coherent
states is nonclassical.

Among pure states, the coherent states are the only ones
satisfying the Glauber classicality criterion
\cite{Hillery-Claspurestatcohe:85}. However, it was recently shown
that coherent states may display nonclassical properties in weak
measurements \cite{Johansen-Noncpropcohestat:03}. This
demonstrates that Glauber's classicality criterion has not taken
into account the possibility of weak measurements. Weak
measurements were discovered by Aharonov, Albert and Vaidman (AAV)
in 1988 \cite{Aharonov+AlbertETAL-ResuMeasCompSpin:88}.

Thermal radiation is almost the only form of naturally occurring
radiation, emanating from any object of finite temperature. It can
be seen in such diverse areas as the cosmic background radiation,
the light from stars and from earthly sources such as a fire or a
light bulb. The $P$-distribution for a thermal state is
\begin{equation}
    P(\alpha) = {1 \over \pi \langle \hat{n} \rangle} e^{- |
    \alpha^2 |/\langle \hat{n} \rangle},
\end{equation}
where $\langle \hat{n} \rangle$ is the expected photon number.
This distribution is neither negative nor singular, hence this is
essentially a classical state according to the Glauber criterion.

In this Letter, we will demonstrate that also thermal radiation
may display nonclassical properties in weak measurements. The term
``nonclassical" means, in this respect, that the effect cannot be
reproduced by a model where the observables are stochastic
$c$-numbers. We will also show that the effect is related to
negativity of the Margenau-Hill quasi-probability distribution
\cite{Margenau+Hill-CorrbetwMeasQuan:61}.

In a standard, projective measurement, the uncertainty of the
pointer is initially very small so that the pointer can
distinguish different eigenvalues of the observable
\cite{Neumann-MathFounQuanMech:55}. In the weak measurement scheme
of AAV, the pointer is assumed to be in a pure, gaussian state of
large uncertainty.  In this way, the pointer cannot discern the
different eigenvalues of the observable. The theory of weak
measurements has been generalized to arbitrary states of object
and pointer in Ref. \cite{Johansen-WeakMeaswithArbi:04}.
Generally, a weak measurement of an observable $\hat{c}$
conditioned on the outcome of projective measurement of an
observable $\hat{d}$ yields the real part of the weak value
\begin{equation}
    c_w(d) = {\langle d \mid \hat{\rho} \hat{c} \mid d \rangle
    \over \langle d \mid \hat{\rho} \mid d \rangle }.
\end{equation}
Here $\hat{\rho}$ is the density operator of the object. $c_w$ is
independent of the specific choice of pointer state. The pointer
may be in an arbitrary state, pure or mixed, provided that the
current density vanishes. Also, it is required that the
interaction between the object and pointer should be sufficiently
weak \cite{Johansen-WeakMeaswithArbi:04}.

The Hamiltonian of the free radiation field is defined as
\begin{equation}\label{eq:hamiltonian}
    \hat{H} = {1 \over 2} (\hat{p}^2 + \hat{q}^2).
\end{equation}
There are two contributions to the Hamiltonian from each of the
quadratures $\hat{q}$ and $\hat{p}$. In this sense, the energy of
the radiation field consists of one energy contribution from each
quadrature. For a material oscillator, one term is kinetic and the
other is potential energy. We shall consider weak measurements of
the energy contribution from one quadrature postselected on the
other. It was recently shown \cite{Johansen-Noncpropcohestat:03}
that the weak value of $\hat{p}^n$ postselected on $\hat{q}$ is a
conditional moment of the standard ordered distribution
\cite{Mehta-PhasFormDynaCano:64}
\begin{equation}\label{eq:condmoment}
    (p^n)_w(q) = {\int dp \; p^n \; S(q,p) \over \langle q \mid
    \hat{\rho} \mid q \rangle},
\end{equation}
where the standard ordered distribution is defined as
\cite{Mehta-PhasFormDynaCano:64}
\begin{equation}
    S(q,p) = {1 \over 2\pi} \int dy \langle q + y \mid
    \hat{\rho} \mid q \rangle e^{-i y p}.
\end{equation}
The standard ordered distribution is the complex conjugate of the
Kirkwood distribution \cite{Kirkwood-QuanStatAlmoClas:33}.

By using Eq. (\ref{eq:pdensity}), we may express the standard
ordered distribution in terms of the $P$-distribution,
\begin{equation}\label{eq:sdistr}
    S(q,p) = {1 \over 2\pi} \int d^2 \alpha P(\alpha) \int dy
    \langle q+y \mid \alpha \rangle \langle \alpha \mid q
    \rangle e^{-i y p}.
\end{equation}
By using the quadrature representation of a coherent state $\mid
\alpha \rangle$ \cite{Louisell-QuanStatPropRadi:73}
\begin{equation}
    \langle q \mid \alpha \rangle = \pi^{-1/4} \exp \left [ -{q^2
    \over 2} + \sqrt{2} \, \alpha \, q - {1 \over 2} \mid \alpha \mid^2
    - {1 \over 2} \, \alpha^2 \right ],
\end{equation}
and by inserting the $P$-distribution for a thermal state in Eq.
(\ref{eq:sdistr}), it can be shown that the standard ordered
distribution for a thermal state is
\begin{equation}\label{eq:standard}
    S(q,p) ={\exp \left [{- 2 \sigma^2 (p^2+q^2) + 2 i p q
    \over {1  + 4 \sigma^4}} \right ] \over \pi \sqrt{1 + 4 \sigma^4}},
\end{equation}
where
\begin{equation}
    \sigma^2 = \langle \hat{n} \rangle + {1 \over 2}
\end{equation}
is the variance of each quadrature.

\begin{figure}

\includegraphics[width=6cm]{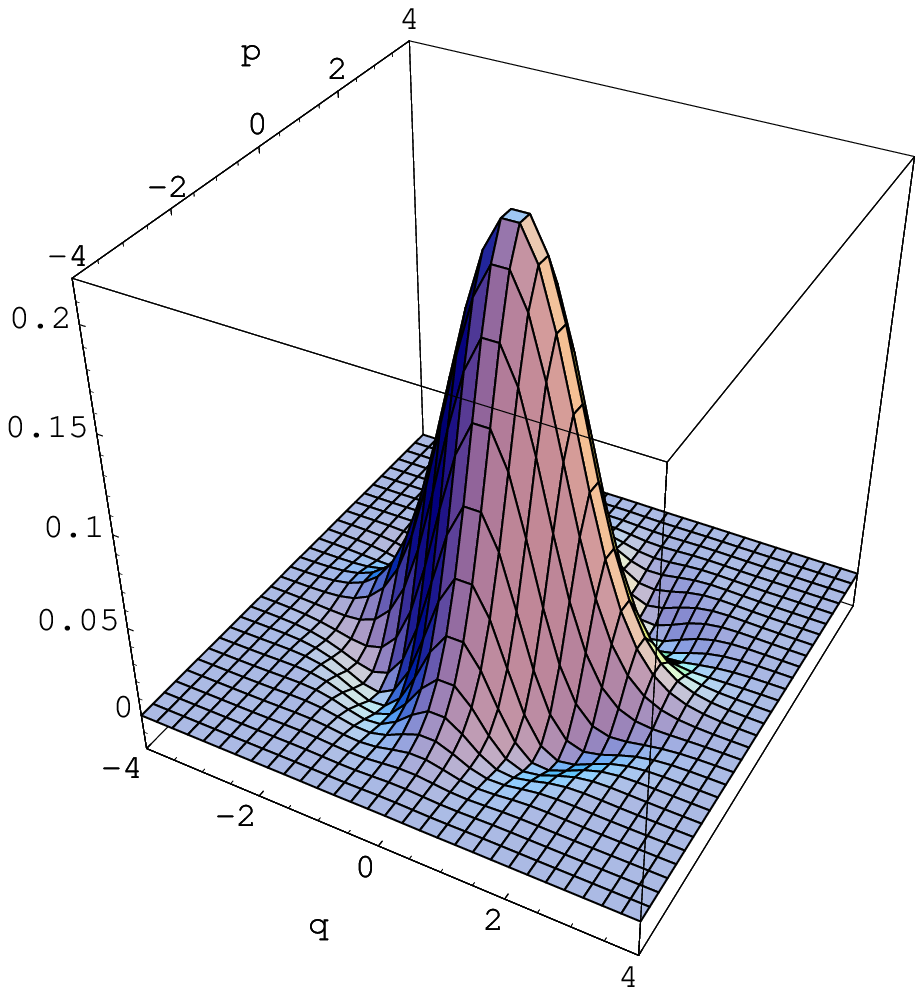}

\caption{The Margenau-Hill distribution for thermal radiation at
maximum irradiance given by Wien's displacement law, i.e., for an
occupation number of $\langle \hat{n}\rangle \approx 10^{-2}$. The
distribution is virtually identical to that of vacuum
\cite{Praxmeyer+Wodkiewicz-QuanInteKirkrepr:03,%
Johansen-Noncpropcohestat:03}. There are distinctly negative
regions, which lead to nonclassical behavior.} \label{fig:mh0}

\includegraphics[width=6cm]{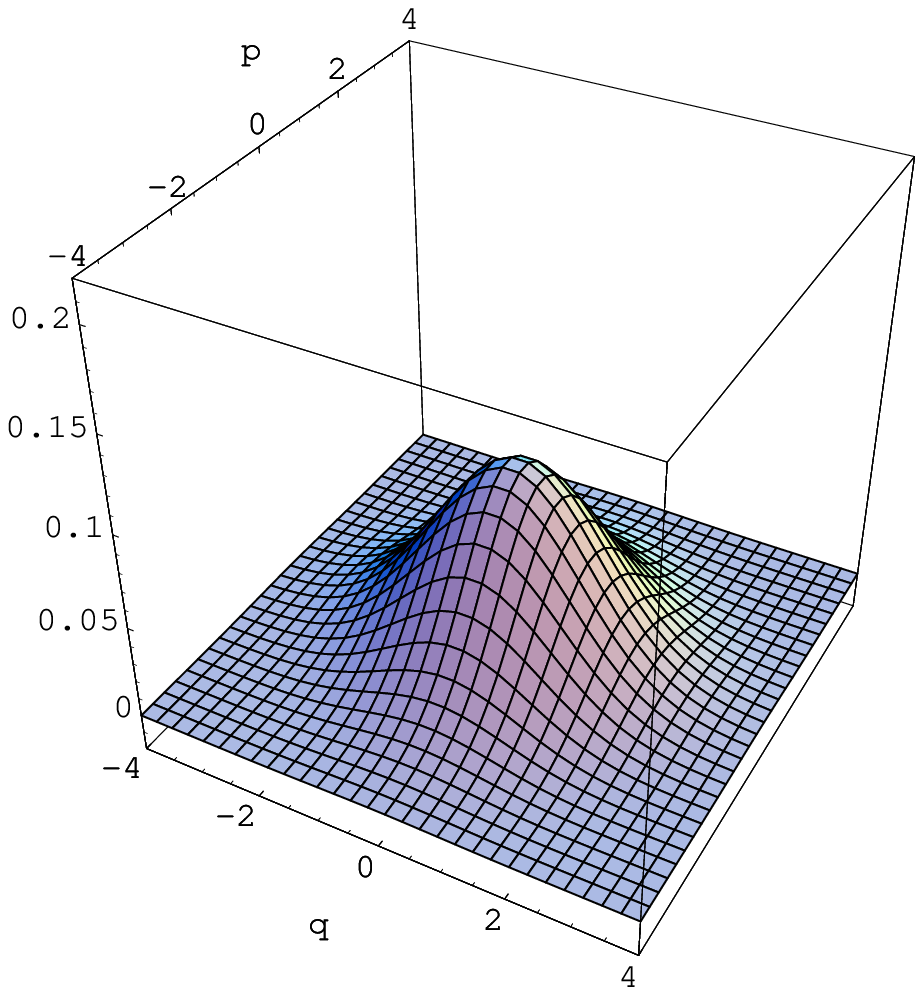}
\caption{The Margenau-Hill distribution for thermal radiation for
an occupation number of $\langle \hat{n}\rangle = 1$. The negative
regions have practically disappeared.} \label{fig:mh1}

\end{figure}

Thermal radiation is often well described by the spectral
distribution of blackbody radiation. For blackbody radiation the
expected occupation number is
\begin{equation}
    \langle \hat{n} \rangle = {1 \over \exp \left ( {\hbar \omega \over k
    T} \right ) - 1}.
\end{equation}
At maximal irradiance of the Planck distribution, as given by
Wien's displacement law, the occupation number is of the order
$10^{-2}$. The Margenau-Hill distribution
\cite{Margenau+Hill-CorrbetwMeasQuan:61}, which is the real part
of the standard ordered distribution, has been plotted in Fig.
\ref{fig:mh0} for an occupation number of $10^{-2}$, and in Fig.
\ref{fig:mh1} at an occupation number of 1.

The standard ordered distribution yields correct marginals when
integrated over. Thus, the marginal distribution for the
quadrature $\hat{q}$ is
\begin{equation}
    \langle q \mid \hat{\rho} \mid q \rangle = \int dp \; S(q,p) =
    {1 \over \sqrt{2 \pi} \sigma} \; e^{-q^2 /(2 \sigma^2)}.
\end{equation}
This is a standard gaussian distribution with variance $\sigma^2$.

By inserting Eq. (\ref{eq:standard}) in Eq. (\ref{eq:condmoment})
we find that the weak value of $\hat{p}^2$ is
\begin{equation}
    (p^2)_w(q) = {\sigma^2 + 4 \sigma^6 - q^2 \over 4 \sigma^4}.
\end{equation}
This is an inverted parabolic curve which is negative for $| q |
\ge \sqrt{\sigma^2 + 4 \sigma^6}$. Thus, for sufficiently large
$q$, the weak value of $\hat{p}^2$ is negative for any thermal
state. A classical radiation model cannot reproduce this result.
The theory of weak measurements can also be extended to classical
theories. It can be shown that if observables are treated as
classical $c$-numbers, the weak value of an observable is just the
conditional expectation value over a positive joint probability
distribution \cite{Johansen-WeakMeaswithArbi:04}. Therefore, in a
classical model, the weak value of a positive observable is always
positive.

\begin{figure}
\includegraphics[width=6cm]{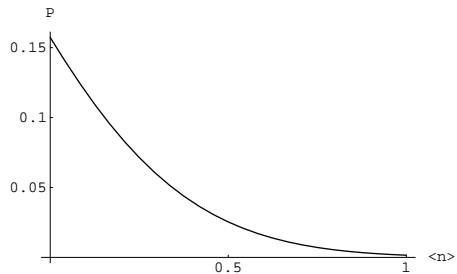}
\caption{The probability of observing a negative weak value of
$\hat{p}^2$ for thermal radiation as a function of the occupation
number $\langle \hat{n}\rangle$. At maximum irradiance of
blackbody radiation, given by Wien's displacement law, the
occupation number is of the order $10^{-2}$, which gives a
significant probability of nonclassicality.} \label{fig:prob}
\end{figure}

The probability of observing a negative weak value is
\begin{equation}
    P = \int_{-\infty}^{-\sqrt{\sigma^2 + 4 \sigma^6}} dq \;
    \langle q \mid \hat{\rho} \mid q \rangle +
    \int_{\sqrt{\sigma^2 + 4 \sigma^6}}^\infty dq \;
    \langle q \mid \hat{\rho} \mid q \rangle.
\end{equation}
The result is
\begin{equation}
    P=\mathrm{erfc} \; \sqrt{{1 \over 2} + 2 \sigma^4}.
\end{equation}
This function has been plotted as a function of $\langle \hat{n}
\rangle$ in Fig. \ref{fig:prob}. We see that the result is
nonclassical essentially for $\langle \hat{n} \rangle < 1$. For
occupation numbers typical of thermal radiation states there is a
significant probability for observing a nonclassical negative weak
value of $\hat{p}^2$.

In principle, weak measurements employ the same interaction
Hamiltonians as strong (or projective) measurements (to this end,
compare Refs. \cite{Johansen-WeakMeaswithArbi:04} and
\cite{Haake+Walls-Overamplmetequan:87}). If a strong measurement
can be performed with a specific interaction Hamiltonian, then a
weak measurement can be performed with the same interaction type
provided that the interaction strength is sufficiently weak.

It could also be possible to observe the nonclassical effect found
in this Letter for a massive particle in thermal equilibrium in a
harmonic oscillator potential. If a detector is placed in a
distance exceeding $\sqrt{\sigma^2 + 4 \sigma^6}$ from it's
equilibrium position, a weak measurement of kinetic energy should
yield a negative value on average. However, this requires
observation in the low temperature regime. For a temperature of 1
$\mu K$, the required oscillator frequency is of the order 100
kHz.

From Eq. (\ref{eq:hamiltonian}) for the Hamiltonian, it can be
shown that
\begin{equation}
    (p^2)_w (q) = 2 {\langle q \mid \hat{\rho} \hat{H} \mid q
    \rangle \over \langle q \mid \hat{\rho} \mid q \rangle } - q^2.
\end{equation}
This suggests an alternative strategy for measuring the weak value
of $\hat{p}^2$. It could also be done by performing a weak
measurement of energy postselected on the quadrature $\hat{q}$,
and subsequently subtracting the squared quadrature \cite{Luis}.

A comment can be made on the usage of pointer states. It has
recently been shown that weak measurements can be performed with
arbitrary pointer states, with the only restriction that the
current density of the pointer state must vanish
\cite{Johansen-WeakMeaswithArbi:04}. This means that a weak
interaction between a thermal pointer state and another thermal
object state may yield a pointer reading which cannot be explained
in terms of a classical $c$-number model of radiation!

In conclusion, we have demonstrated that thermal radiation
displays strong nonclassicality in weak measurements. The effect
is present for low occupation numbers, as found in most naturally
occurring forms of radiation. A weak measurement of the squared
quadrature postselected on the second quadrature may yield a
negative weak value. This effect cannot be reproduced by a
classical model of radiation where observables are $c$-numbers.
The nonclassical effect is related to negativity of the
Margenau-Hill distribution for the thermal state.


\begin{thebibliography}{18}
\bibitem[{\citenamefont{Luis}()}]{Luis}
  \bibinfo{howpublished}{Alfredo Luis is currently studying the
  implementation of weak energy measurements for coherent and
  thermal states with the author}.
\end{thebibliography}
\end{document}